\documentclass[%
superscriptaddress,
preprint,
 amsmath,amssymb,
 aps,
apl,
]{revtex4-1}

\usepackage{graphicx}
\usepackage{dcolumn}
\usepackage{bm}
\usepackage{amsmath}
\usepackage{epsfig}
\usepackage{rotating}
\usepackage{enumerate}
\usepackage{xcolor}
\usepackage{color,soul}
\usepackage{siunitx}

\begin{document}

\title{Micro-electromechanical memory bit based on magnetic repulsion}

\author{Miquel L\'opez-Su\'arez}
\affiliation{NiPS Laboratory, Dipartimento di Fisica e Geologia,
             Universit\`a degli Studi di Perugia, 
             06123 Perugia, Italy}

\author{Igor Neri}
\affiliation{NiPS Laboratory, Dipartimento di Fisica e Geologia,
             Universit\`a degli Studi di Perugia, 
             06123 Perugia, Italy}
\affiliation{INFN Sezione di Perugia, via Pascoli,
             06123 Perugia, Italy}
\email{igor.neri@nipslab.org}

\date{\today}

\begin{abstract}
A bistable micro-mechanical system based on magnetic repulsion is presented exploring its applicability as memory unit where the state of the bit is encoded in the rest position of a deflected cantilever. The non-linearity induced on the cantilever can be tuned through the magnetic interaction intensity between the cantilever magnet and the counter magnet in terms of geometrical parameters. A simple model provides a sound prediction of the behavior of the system. Finally we measured the energy required to store a bit of information on the system that, for the considered protocols, is bounded by the energy barrier separating the two stable states.
\end{abstract}

\maketitle

\begin{figure}
\includegraphics[width=0.75\textwidth]{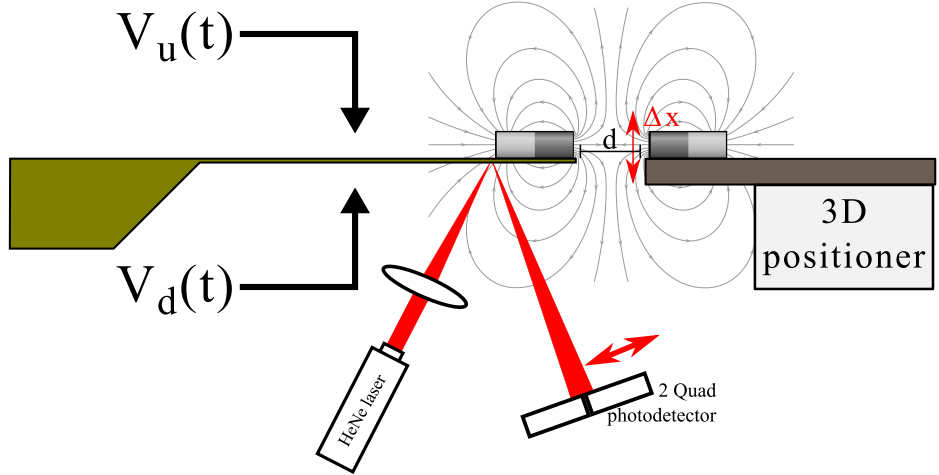}
\caption{Schematic of the whole system and measurement setup, lateal view. The bending of the cantilever ($k$=\SI{0.08}{\newton\per\meter}, $f_0$=\SI{5.3}{\kilo\Hz}) is measured by means of an optical lever. Two magnets with opposite magnetic orientations are used to introduce bistability. Two electrodes are used to apply electrostatic forces on the mechanical structure: $V_u$ and $V_d$ to force the cantilever to bend upwards and downwards respectively. The magnetic interaction can be engineered by changing geometric parameters such as $d$ and $\Delta x$.}\label{f:schematic}
\end{figure}

Mechanical bistability has attracted the attention of the scientific community in the last years as the base for the implementation of switches \cite{martini2016experimental,madami2015fundamental}, memory bits \cite{roodenburg2009buckling}, vibration energy harvesters \cite{cottone2009nonlinear} and sensitive measurement systems \cite{badzey2005coherent}. In previous works we have introduced and demonstrated the feasibility of inducing bistable behavior in Micro Electro Mechanical Systems (MEMS) through electrostatic interaction \cite{lopez2013inducing} and mechanical compression \cite{lopez2014piezoelectric,neri2015reset}. However the implementation and study of new routes to achieve bistable behavior at the micro and nanoscale are still of technologically and scientific interest. In this work we present a bistable system implemented with a micro-cantilever with a permanent magnet attached to its tip. Bistability is introduced by placing a counter magnet, with opposite magnetization, in front of the cantilever as depicted in Fig.~\ref{f:schematic}. The magnetic interaction can be controlled by varying the distance between the two magnets, $d$, or changing their relative alignment, $\Delta x$. The behavior of the system is characterized in terms of the rest position of the cantilever tip, $x_{eq}$, and resonant frequency, $f_0$. We present a model able to capture the mean features of the system as a function of the effective magnetization and geometric parameters. Finally, a realization of a micro-mechanical memory bit is demonstrated, estimating the energy dissipated to store one bit of information on the system.

A commercial triangular micro-cantilever (NanoWorld PNP-TR-TL\cite{nanoworld}) is used as mechanical structure. The cantilever is long \SI{200}{\micro\meter}, with a nominal stiffness of $k$=\SI{0.08}{\newton\per\meter}. A fragment of NdFeB (neodymium) magnet is attached to the cantilever tip with bi-component epoxy resin. To set the magnetization to a known direction the system is heated up to \SI{670}{\kelvin}, above its Curie temperature \cite{ma2002recent}, in the presence of a strong external magnetic field with the desired orientation. A set of two electrodes (see Fig.~\ref{f:schematic}) is used in order to polarize the cantilever producing a bend on the mechanical structure. The electrostatic force depends on the voltage applied to the electrodes, i.e.\ $V_u$ and $V_d$.

The deflection of the cantilever, $x$, is determined by means of a AFM-like laser optical lever \cite{alexander1989atomic} where $x$ is proportional to the output voltage provided by the two quadrants photodetector (see Fig.~\ref{f:schematic}). The system is placed in a vacuum chamber  and isolated from seismic vibrations to maximize the signal-to-noise ratio. All measurements were performed at pressure $P$=\SI{4.7e-2}{\milli\bar}. The counter magnet position is set with a 3D stage in order to find the working position. Two piezoelectric stacks, with a maximum excursion of \SI{20}{\micro\meter}, are then used for a fine control of $\Delta x$, and $d$.

\begin{figure}
\includegraphics[width=0.75\textwidth]{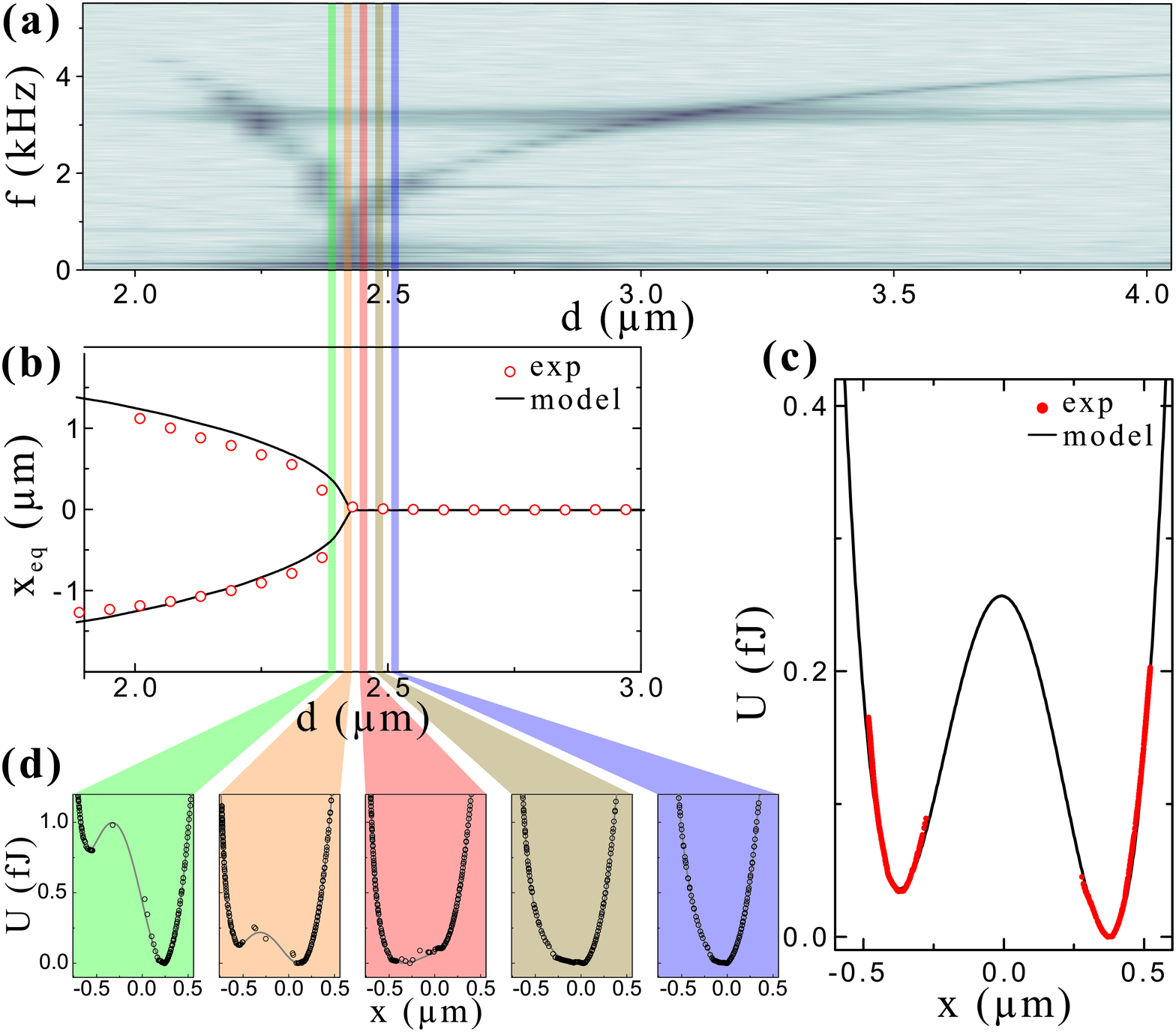}
\caption{Frequency response and bistability of the system as a function of the distance between magnets. (a) Frequency response of the cantilever as a function of the distance between magnets, $d$. Reducing the distance between magnets the magnetic repulsion increases. The effect is a softening of the structure and thus a lowering of its resonant frequency. At the bifurcation point, $d_b$, the resonant frequency reaches a minimum value lower than \SI{500}{\Hz}. For $d<d_b$, $f_0$ increases continuously. (b) Map of the rest positon of the cantilever as a function of $d$. A bifurcation appears at $d=d_b$ where bistability is generated. A good agreement between model (black line) and experiment (red circles) is achieved. (c) Total potential energy, $U$, for $d$=\SI{2.392}{\micro\meter}. $\Delta x$=\SI{6}{\angstrom} is introduced into the model (black line) in order to reproduce the tilt of \SI{0.04}{\femto\joule} shown by the experimental data (red circles). (d) Total potential energy, $U$, for different values of the controlling parameter, $d$, showing the flattening of the energy landscape and the appearing of bistability. From right to left: $d$=\SI{2.535}{\micro\meter} blue panel; $d$=\SI{2.505}{\micro\meter} ocre panel; $d$=\SI{2.475}{\micro\meter} red panel; $d$=\SI{2.445}{\micro\meter} orange panel; $d$=\SI{2.415}{\micro\meter} green panel.}\label{f:bifurcation}
\end{figure}

The frequency response of the cantilever is shown in Fig.~\ref{f:bifurcation}(a) as a function of $d$. Decreasing $d$, the mechanical structure softens producing a shift in its resonant frequency, $f_0$, towards lower values. This reaches a minimum value for $d=d_b$ (where $d_b$ is the critical distance at which bistability appears) lower than \SI{500}{\Hz} corresponding to a decrease of approximately \SI{99}{\percent} respect its natural resonant frequency, $f_0$=\SI{5.3}{\kilo\Hz} measured without magnetic interaction). 

Mechanical structures like cantilevers are usually modeled as linear systems with a elastic potential energy $U_{el}=1/2kx^2$, where $k$ is the elastic constant. In order to introduce into the model the magnetic interaction we take into account the force between two magnetic dipoles as:
\begin{multline*}
\mathbf{F_m}(\mathbf{r}, \mathbf{m}_1, \mathbf{m}_2) = \frac{3 \mu_0}{4 \pi r^5}\bigl[(\mathbf{m}_1\cdot\mathbf{r})\mathbf{m}_2 + (\mathbf{m}_2\cdot\mathbf{r})\mathbf{m}_1 
 + (\mathbf{m}_1\cdot\mathbf{m}_2)\mathbf{r} - \frac{5(\mathbf{m}_1\cdot\mathbf{r})(\mathbf{m}_2\cdot\mathbf{r})}{r^2}\mathbf{r}\bigr]
\end{multline*}

where $\mathbf{m}_1$ and $\mathbf{m}_2$ are the dipole magnetizations, $\mathbf{r}$ is the distance between them, and $\mu_0$ is the vacuum permeability, which yields to the corresponding magnetic potential energy, $U_m=-\int \! F_m  \, \mathrm{d}x$. The rest position of the cantilever, $x_{eq}$, is obtained by finding the minima of the total potential energy $U=U_m+U_{el}$. In Fig.~\ref{f:bifurcation}(b) a comparison of $x_{eq}$ predicted by the model, without asymmetries(black line), and measured experimentally (red circles) as a function of $d$ is shown. The effective magnetization, $m^*=\sqrt{\mathbf{m}_1 \cdot\mathbf{m}_2}$, is set to match the distance $d_b$ at which bifurcation of $x_{eq}$ occurs. The black line in Fig.~\ref{f:bifurcation}(c) shows the modeled potential energy for $d$=\SI{2.392}{\micro\meter}. The total potential energy $U$, has been reconstructed applying a set of known forces through the electrodes and measuring the rest position of the cantilever. The potential has been computed evaluating $U=\int \! F_{el} \, \mathrm{d}x$ numerically, where $F_{el}$ is the electrostatic force produced by the electrodes.
A lateral misalignment of $\Delta x$=\SI{6}{\angstrom} has been set to fit the tilt of the measured potential energy. The two wells show an energy difference of \SI{0.04}{\femto\joule}. The model (black line) shows a good agreement with the experimental values (red circles).
Panels in Fig.~\ref{f:bifurcation}(d) show the potential energies of the system for the values of $d$ highlighted by colored regions. The system is monostable for $d>d_b$, although a flattering of $U$ can be seen. At some point between the red and orange regions the bifurcation of $x_{eq}$ occurs. For $d<d_b$ two stable states are present (orange and green regions).

\begin{figure}
\includegraphics[width=0.75\textwidth]{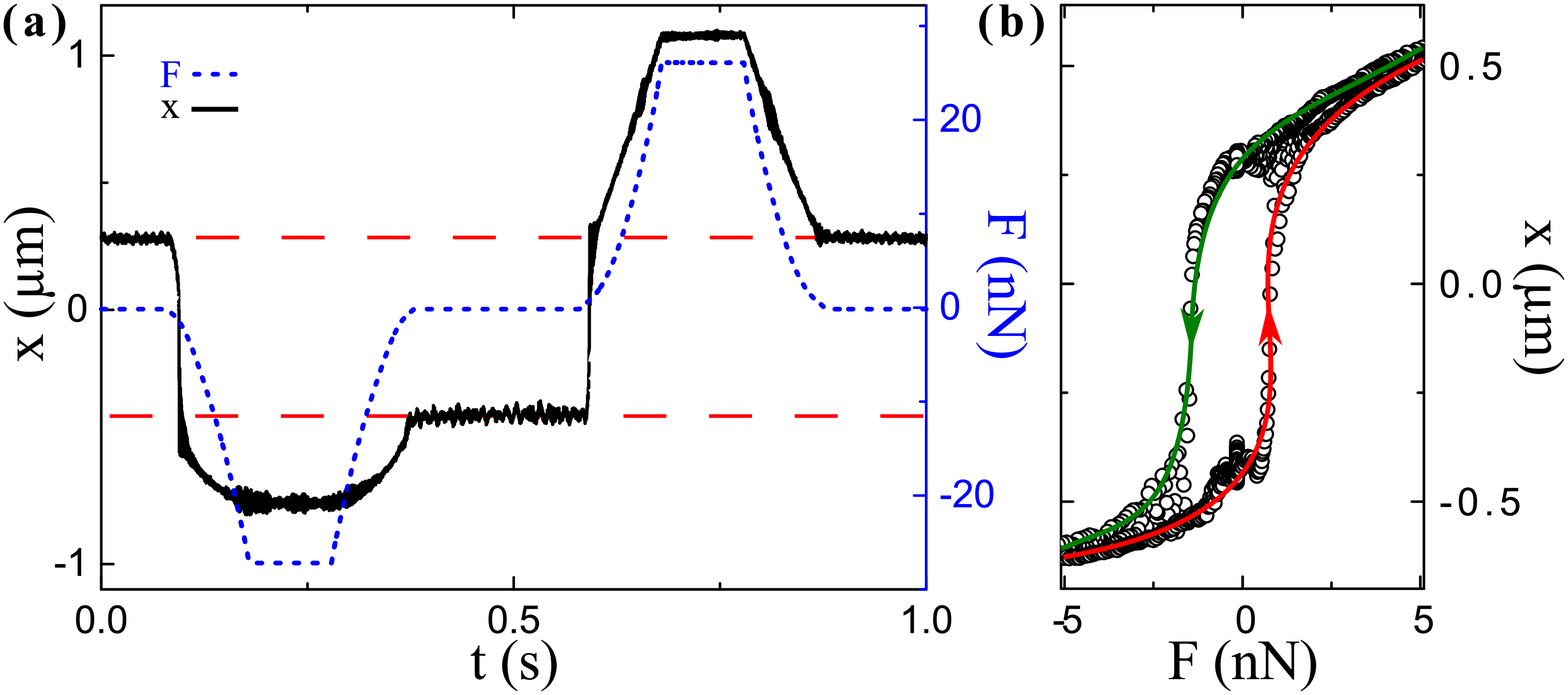}
\caption{Mechanical responce of the system to electrostatic forces. (a) Time series of the displacement of the cantilever tip (black solid curve) relative to the applied eletrostatic forces (blue dotted curve) when the system is bistable. The two rest positions correspond to the dashed red curves. When the force is zero the cantilever is in one of the rest position. (b) Hysteresis loop of the position as function of the applied force, $F$. The superimposed red curve shows the path followed by the cantilever position when the forces goes from negative to positive, starting from the state `0'; the green curve shows the cantilever tip path when the force goes from positive to negative, starting from the state `1'.}\label{f:hysteresis}
\end{figure}

As pointed out in the introduction of this paper, the implementation of bistable systems at the microscale is pursued in many fields. In the following we show the application of the system presented below as a micromechanical memory bit. We encode the information in the position of the cantilever tip, i.e.\ `0' for $x<0$ and `1' for $x>0$.
The information stored in the system is changed applying a set of electrostatic forces through the two electrodes. In the case presented in Fig.~\ref{f:hysteresis}(a) the forces, represented in as blue dotted curve, are applied cyclically to switch from `1' to `0' and vice versa. Those forces result in an effective tilt of the potential energy that push the cantilever to jump from one to the other potential well (black solid curve). Once the voltages, and thus the forces, are set to zero the state of the bit can be read. The two rest positions are represented by the dashed red curves.
Fig.~\ref{f:hysteresis}(b) shows the hysteresis loop of $x$ as a function of the electrostatic forces $F$. The superimposed red curve shows the path followed by the cantilever position when increasing the force and going positive, starting from the state `0', the green curve shows the cantilever tip path when the forces goes negative, starting from the state `1'.

\begin{figure}
\includegraphics[width=0.75\textwidth]{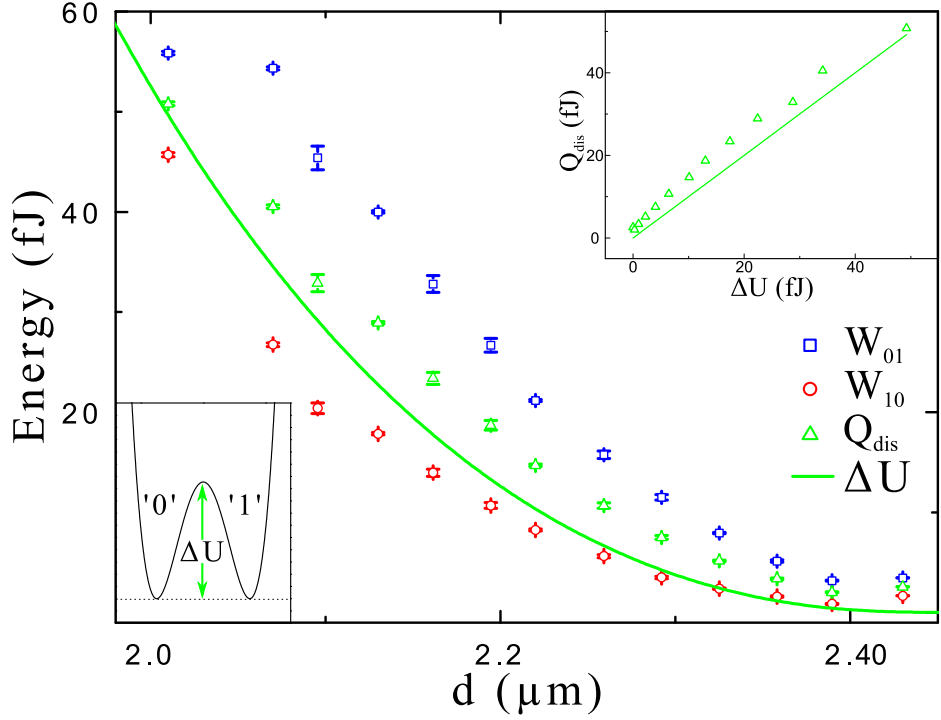}
\caption{Energy dissipation, and work done on the system, switching the system from state `0' to `1' and vice versa for different magnets distance, $d$. Blue squares represent the average work done on the system to switch from state `0' to `1'. Red circles represent the average work done on the system to switch from state `1' to `0'. Green triangules represent the average energy disspated, computed averaging between $W_{01}$ and $W_{10}$. Green solid line represents the energy barrier separating the two stable states for a symmetric system, as function of magnets distance. Top right inset shows the dependance between $\Delta U$ and $Q_{dis}$, where the solid line represent the bisector of the graph, and thus the minimum energy to be dissipated.}\label{f:dissipation}
\end{figure}

Energy dissipation per logic operation is one of the quantities that must be addressed by future logic devices to further increase their performance. We have measured the work per switch done on the system by evaluating $W=\int \! \frac{\partial U}{\partial V} \, \mathrm{d}V$ as in \cite{jun2014high,berut2012experimental,lopez2015operating}.
Red circles and blue squares in Fig.~\ref{f:dissipation} represent the work performed to switch the bit from `0' to `1' ($W_{01}$) and from `1' to `0' ($W_{10}$) for different values of $d$. The difference between $W_{01}$ and $W_{10}$ is due to the potential energy difference in the bottom of the two potential wells. 

To evaluate the average energy dissipated per bit stored, $Q_{dis}$, both transitions, `0' to `1' and `1' to `0', have to be taken into account. On average thus the potential energy change is equal to zero and then the dissipated energy is the average of the work done for the two transitions: $Q_{dis}=\frac{W_{01}+W_{10}}{2}$. The measured values for $Q_{dis}$ are shown in Fig.~\ref{f:dissipation} with green triangles, showing a clear decrease of energy dissipated as $d$ increases. Notice that while $d$ increases the energy barrier ,$\Delta U$, separating the two wells decreases, and the two minima are less separated. As a matter of fact the forces applied to switch the bit are used to bring the system on the top of the energy barrier separating the two stable states. After reaching this point the system transits to the other well and must dissipate the difference between the potential energy and the thermal energy of the system, this is at least $\Delta U - k_BT$. In the present case $\Delta U$ is much larger than $k_BT$, therefore we can say that $\Delta U$ sets the minimum energy dissipation for the present system, represented with a green line in Fig.~\ref{f:dissipation}. As expected, $Q_{dis}$ is above $\Delta U$ for all the considered $d$ values. The dependance between $\Delta U$ and $Q_{dis}$ is more clear in the top right inset of Fig.~\ref{f:dissipation}, where the solid line represent the bisector of the graph, and thus the minimum energy to be dissipated using this protocol.
Notice that the switch can be achieved in a more energetically convinient way using ad hoc protocols, e.g., removing the barrier \cite{madami2015fundamental,neri2015reset}, in order to reduce the dissipation to its physical limit, i.e., arbitrarly low energy dissipated. However intrinsic damping mechanisms (e.g. structural damping) can set a higher limit to the energy disspated depending on the maximum deflection of the cantilever\cite{lopez2015sub}. A downscale of the system dimensions could help in order to reduce energy dissipation. However, this will cause a dramatic increase of its elastic constant, $k$. In addition downscaling the magnet dimensions will also reduce its total magnetization. As a consequence of both effects bistability would be hard to obtain and a smart route to bring this sort of device to the nanoscale is needed.

In summary, we have modeled, fabricated and tested a novel implementation of a bistable system at the microscale where magnetic repulsion is used to engineer the non-linearity. The distance between magnets and the horizontal alignment are used to tune the intensity of the magnetic interaction. We have also shown that the system can be operated as a micro-mechanical memory bit, measuring a minimum energy dissipation produced, storing a  bit of information, lower than \SI{1}{\femto\joule}.

\begin{acknowledgments}
Authors gratefully acknowledge financial support from the European 
Commission (FPVII, Grant agreement no: 318287, LANDAUER and Grant 
agreement no: 611004, ICT-Energy). The authors thank L. Gammaitoni
for useful discussions.
\end{acknowledgments}

\bibliography{bibliography}

\begin{thebibliography}{15}%
\makeatletter
\providecommand \@ifxundefined [1]{%
 \@ifx{#1\undefined}
}%
\providecommand \@ifnum [1]{%
 \ifnum #1\expandafter \@firstoftwo
 \else \expandafter \@secondoftwo
 \fi
}%
\providecommand \@ifx [1]{%
 \ifx #1\expandafter \@firstoftwo
 \else \expandafter \@secondoftwo
 \fi
}%
\providecommand \natexlab [1]{#1}%
\providecommand \enquote  [1]{``#1''}%
\providecommand \bibnamefont  [1]{#1}%
\providecommand \bibfnamefont [1]{#1}%
\providecommand \citenamefont [1]{#1}%
\providecommand \href@noop [0]{\@secondoftwo}%
\providecommand \href [0]{\begingroup \@sanitize@url \@href}%
\providecommand \@href[1]{\@@startlink{#1}\@@href}%
\providecommand \@@href[1]{\endgroup#1\@@endlink}%
\providecommand \@sanitize@url [0]{\catcode `\\12\catcode `\$12\catcode
  `\&12\catcode `\#12\catcode `\^12\catcode `\_12\catcode `\%12\relax}%
\providecommand \@@startlink[1]{}%
\providecommand \@@endlink[0]{}%
\providecommand \url  [0]{\begingroup\@sanitize@url \@url }%
\providecommand \@url [1]{\endgroup\@href {#1}{\urlprefix }}%
\providecommand \urlprefix  [0]{URL }%
\providecommand \Eprint [0]{\href }%
\providecommand \doibase [0]{http://dx.doi.org/}%
\providecommand \selectlanguage [0]{\@gobble}%
\providecommand \bibinfo  [0]{\@secondoftwo}%
\providecommand \bibfield  [0]{\@secondoftwo}%
\providecommand \translation [1]{[#1]}%
\providecommand \BibitemOpen [0]{}%
\providecommand \bibitemStop [0]{}%
\providecommand \bibitemNoStop [0]{.\EOS\space}%
\providecommand \EOS [0]{\spacefactor3000\relax}%
\providecommand \BibitemShut  [1]{\csname bibitem#1\endcsname}%
\let\auto@bib@innerbib\@empty
\bibitem [{\citenamefont {Martini}\ \emph {et~al.}(2016)\citenamefont
  {Martini}, \citenamefont {Pancaldi}, \citenamefont {Madami}, \citenamefont
  {Vavassori}, \citenamefont {Gubbiotti}, \citenamefont {Tacchi}, \citenamefont
  {Hartmann}, \citenamefont {Emmerling}, \citenamefont {H{\"o}fling},
  \citenamefont {Worschech} \emph {et~al.}}]{martini2016experimental}%
  \BibitemOpen
  \bibfield  {author} {\bibinfo {author} {\bibfnamefont {L.}~\bibnamefont
  {Martini}}, \bibinfo {author} {\bibfnamefont {M.}~\bibnamefont {Pancaldi}},
  \bibinfo {author} {\bibfnamefont {M.}~\bibnamefont {Madami}}, \bibinfo
  {author} {\bibfnamefont {P.}~\bibnamefont {Vavassori}}, \bibinfo {author}
  {\bibfnamefont {G.}~\bibnamefont {Gubbiotti}}, \bibinfo {author}
  {\bibfnamefont {S.}~\bibnamefont {Tacchi}}, \bibinfo {author} {\bibfnamefont
  {F.}~\bibnamefont {Hartmann}}, \bibinfo {author} {\bibfnamefont
  {M.}~\bibnamefont {Emmerling}}, \bibinfo {author} {\bibfnamefont
  {S.}~\bibnamefont {H{\"o}fling}}, \bibinfo {author} {\bibfnamefont
  {L.}~\bibnamefont {Worschech}},  \emph {et~al.},\ }\href@noop {} {\bibfield
  {journal} {\bibinfo  {journal} {Nano Energy}\ }\textbf {\bibinfo {volume}
  {19}},\ \bibinfo {pages} {108} (\bibinfo {year} {2016})}\BibitemShut
  {NoStop}%
\bibitem [{\citenamefont {Madami}\ \emph {et~al.}(2015)\citenamefont {Madami},
  \citenamefont {Chiuchi{\`u}}, \citenamefont {Carlotti},\ and\ \citenamefont
  {Gammaitoni}}]{madami2015fundamental}%
  \BibitemOpen
  \bibfield  {author} {\bibinfo {author} {\bibfnamefont {M.}~\bibnamefont
  {Madami}}, \bibinfo {author} {\bibfnamefont {D.}~\bibnamefont
  {Chiuchi{\`u}}}, \bibinfo {author} {\bibfnamefont {G.}~\bibnamefont
  {Carlotti}}, \ and\ \bibinfo {author} {\bibfnamefont {L.}~\bibnamefont
  {Gammaitoni}},\ }\href@noop {} {\bibfield  {journal} {\bibinfo  {journal}
  {Nano Energy}\ }\textbf {\bibinfo {volume} {15}},\ \bibinfo {pages} {313}
  (\bibinfo {year} {2015})}\BibitemShut {NoStop}%
\bibitem [{\citenamefont {Roodenburg}\ \emph {et~al.}(2009)\citenamefont
  {Roodenburg}, \citenamefont {Spronck}, \citenamefont {Van~der Zant},\ and\
  \citenamefont {Venstra}}]{roodenburg2009buckling}%
  \BibitemOpen
  \bibfield  {author} {\bibinfo {author} {\bibfnamefont {D.}~\bibnamefont
  {Roodenburg}}, \bibinfo {author} {\bibfnamefont {J.}~\bibnamefont {Spronck}},
  \bibinfo {author} {\bibfnamefont {H.}~\bibnamefont {Van~der Zant}}, \ and\
  \bibinfo {author} {\bibfnamefont {W.}~\bibnamefont {Venstra}},\ }\href@noop
  {} {\bibfield  {journal} {\bibinfo  {journal} {Applied Physics Letters}\
  }\textbf {\bibinfo {volume} {94}},\ \bibinfo {pages} {183501} (\bibinfo
  {year} {2009})}\BibitemShut {NoStop}%
\bibitem [{\citenamefont {Cottone}\ \emph {et~al.}(2009)\citenamefont
  {Cottone}, \citenamefont {Vocca},\ and\ \citenamefont
  {Gammaitoni}}]{cottone2009nonlinear}%
  \BibitemOpen
  \bibfield  {author} {\bibinfo {author} {\bibfnamefont {F.}~\bibnamefont
  {Cottone}}, \bibinfo {author} {\bibfnamefont {H.}~\bibnamefont {Vocca}}, \
  and\ \bibinfo {author} {\bibfnamefont {L.}~\bibnamefont {Gammaitoni}},\
  }\href@noop {} {\bibfield  {journal} {\bibinfo  {journal} {Physical Review
  Letters}\ }\textbf {\bibinfo {volume} {102}},\ \bibinfo {pages} {080601}
  (\bibinfo {year} {2009})}\BibitemShut {NoStop}%
\bibitem [{\citenamefont {Badzey}\ and\ \citenamefont
  {Mohanty}(2005)}]{badzey2005coherent}%
  \BibitemOpen
  \bibfield  {author} {\bibinfo {author} {\bibfnamefont {R.~L.}\ \bibnamefont
  {Badzey}}\ and\ \bibinfo {author} {\bibfnamefont {P.}~\bibnamefont
  {Mohanty}},\ }\href@noop {} {\bibfield  {journal} {\bibinfo  {journal}
  {Nature}\ }\textbf {\bibinfo {volume} {437}},\ \bibinfo {pages} {995}
  (\bibinfo {year} {2005})}\BibitemShut {NoStop}%
\bibitem [{\citenamefont {L{\'o}pez-Su{\'a}rez}\ \emph
  {et~al.}(2013)\citenamefont {L{\'o}pez-Su{\'a}rez}, \citenamefont {Agusti},
  \citenamefont {Torres}, \citenamefont {Rurali},\ and\ \citenamefont
  {Abadal}}]{lopez2013inducing}%
  \BibitemOpen
  \bibfield  {author} {\bibinfo {author} {\bibfnamefont {M.}~\bibnamefont
  {L{\'o}pez-Su{\'a}rez}}, \bibinfo {author} {\bibfnamefont {J.}~\bibnamefont
  {Agusti}}, \bibinfo {author} {\bibfnamefont {F.}~\bibnamefont {Torres}},
  \bibinfo {author} {\bibfnamefont {R.}~\bibnamefont {Rurali}}, \ and\ \bibinfo
  {author} {\bibfnamefont {G.}~\bibnamefont {Abadal}},\ }\href@noop {}
  {\bibfield  {journal} {\bibinfo  {journal} {Applied Physics Letters}\
  }\textbf {\bibinfo {volume} {102}},\ \bibinfo {pages} {153901} (\bibinfo
  {year} {2013})}\BibitemShut {NoStop}%
\bibitem [{\citenamefont {L{\'o}pez-Su{\'a}rez}\ \emph
  {et~al.}(2014)\citenamefont {L{\'o}pez-Su{\'a}rez}, \citenamefont {Pruneda},
  \citenamefont {Abadal},\ and\ \citenamefont
  {Rurali}}]{lopez2014piezoelectric}%
  \BibitemOpen
  \bibfield  {author} {\bibinfo {author} {\bibfnamefont {M.}~\bibnamefont
  {L{\'o}pez-Su{\'a}rez}}, \bibinfo {author} {\bibfnamefont {M.}~\bibnamefont
  {Pruneda}}, \bibinfo {author} {\bibfnamefont {G.}~\bibnamefont {Abadal}}, \
  and\ \bibinfo {author} {\bibfnamefont {R.}~\bibnamefont {Rurali}},\
  }\href@noop {} {\bibfield  {journal} {\bibinfo  {journal} {Nanotechnology}\
  }\textbf {\bibinfo {volume} {25}},\ \bibinfo {pages} {175401} (\bibinfo
  {year} {2014})}\BibitemShut {NoStop}%
\bibitem [{\citenamefont {Neri}\ \emph {et~al.}(2015)\citenamefont {Neri},
  \citenamefont {Lopez-Suarez}, \citenamefont {Chiuchi{\'u}},\ and\
  \citenamefont {Gammaitoni}}]{neri2015reset}%
  \BibitemOpen
  \bibfield  {author} {\bibinfo {author} {\bibfnamefont {I.}~\bibnamefont
  {Neri}}, \bibinfo {author} {\bibfnamefont {M.}~\bibnamefont {Lopez-Suarez}},
  \bibinfo {author} {\bibfnamefont {D.}~\bibnamefont {Chiuchi{\'u}}}, \ and\
  \bibinfo {author} {\bibfnamefont {L.}~\bibnamefont {Gammaitoni}},\
  }\href@noop {} {\bibfield  {journal} {\bibinfo  {journal} {EPL (Europhysics
  Letters)}\ }\textbf {\bibinfo {volume} {111}},\ \bibinfo {pages} {10004}
  (\bibinfo {year} {2015})}\BibitemShut {NoStop}%
\bibitem [{nan()}]{nanoworld}%
  \BibitemOpen
  \href@noop {} {\enquote {\bibinfo {title} {{NanoWorld - AFM tip - PNP-TR-TL -
  Pyrex-Nitride}},}\ }\bibinfo {howpublished}
  {http://www.nanoworld.com/pyrex-nitride-triangular-silicon-nitride-tipless-cantilever-afm-tip-pnp-tr-tl},\
  \bibinfo {note} {accessed: 2016-02-25}\BibitemShut {NoStop}%
\bibitem [{\citenamefont {Ma}\ \emph {et~al.}(2002)\citenamefont {Ma},
  \citenamefont {Herchenroeder}, \citenamefont {Smith}, \citenamefont {Suda},
  \citenamefont {Brown},\ and\ \citenamefont {Chen}}]{ma2002recent}%
  \BibitemOpen
  \bibfield  {author} {\bibinfo {author} {\bibfnamefont {B.}~\bibnamefont
  {Ma}}, \bibinfo {author} {\bibfnamefont {J.}~\bibnamefont {Herchenroeder}},
  \bibinfo {author} {\bibfnamefont {B.}~\bibnamefont {Smith}}, \bibinfo
  {author} {\bibfnamefont {M.}~\bibnamefont {Suda}}, \bibinfo {author}
  {\bibfnamefont {D.}~\bibnamefont {Brown}}, \ and\ \bibinfo {author}
  {\bibfnamefont {Z.}~\bibnamefont {Chen}},\ }\href@noop {} {\bibfield
  {journal} {\bibinfo  {journal} {Journal of magnetism and magnetic materials}\
  }\textbf {\bibinfo {volume} {239}},\ \bibinfo {pages} {418} (\bibinfo {year}
  {2002})}\BibitemShut {NoStop}%
\bibitem [{\citenamefont {Alexander}\ \emph {et~al.}(1989)\citenamefont
  {Alexander}, \citenamefont {Hellemans}, \citenamefont {Marti}, \citenamefont
  {Schneir}, \citenamefont {Elings}, \citenamefont {Hansma}, \citenamefont
  {Longmire},\ and\ \citenamefont {Gurley}}]{alexander1989atomic}%
  \BibitemOpen
  \bibfield  {author} {\bibinfo {author} {\bibfnamefont {S.}~\bibnamefont
  {Alexander}}, \bibinfo {author} {\bibfnamefont {L.}~\bibnamefont
  {Hellemans}}, \bibinfo {author} {\bibfnamefont {O.}~\bibnamefont {Marti}},
  \bibinfo {author} {\bibfnamefont {J.}~\bibnamefont {Schneir}}, \bibinfo
  {author} {\bibfnamefont {V.}~\bibnamefont {Elings}}, \bibinfo {author}
  {\bibfnamefont {P.~K.}\ \bibnamefont {Hansma}}, \bibinfo {author}
  {\bibfnamefont {M.}~\bibnamefont {Longmire}}, \ and\ \bibinfo {author}
  {\bibfnamefont {J.}~\bibnamefont {Gurley}},\ }\href@noop {} {\bibfield
  {journal} {\bibinfo  {journal} {Journal of Applied Physics}\ }\textbf
  {\bibinfo {volume} {65}},\ \bibinfo {pages} {164} (\bibinfo {year}
  {1989})}\BibitemShut {NoStop}%
\bibitem [{\citenamefont {Jun}\ \emph {et~al.}(2014)\citenamefont {Jun},
  \citenamefont {Gavrilov},\ and\ \citenamefont {Bechhoefer}}]{jun2014high}%
  \BibitemOpen
  \bibfield  {author} {\bibinfo {author} {\bibfnamefont {Y.}~\bibnamefont
  {Jun}}, \bibinfo {author} {\bibfnamefont {M.}~\bibnamefont {Gavrilov}}, \
  and\ \bibinfo {author} {\bibfnamefont {J.}~\bibnamefont {Bechhoefer}},\
  }\href@noop {} {\bibfield  {journal} {\bibinfo  {journal} {Physical review
  letters}\ }\textbf {\bibinfo {volume} {113}},\ \bibinfo {pages} {190601}
  (\bibinfo {year} {2014})}\BibitemShut {NoStop}%
\bibitem [{\citenamefont {B{\'e}rut}\ \emph {et~al.}(2012)\citenamefont
  {B{\'e}rut}, \citenamefont {Arakelyan}, \citenamefont {Petrosyan},
  \citenamefont {Ciliberto}, \citenamefont {Dillenschneider},\ and\
  \citenamefont {Lutz}}]{berut2012experimental}%
  \BibitemOpen
  \bibfield  {author} {\bibinfo {author} {\bibfnamefont {A.}~\bibnamefont
  {B{\'e}rut}}, \bibinfo {author} {\bibfnamefont {A.}~\bibnamefont
  {Arakelyan}}, \bibinfo {author} {\bibfnamefont {A.}~\bibnamefont
  {Petrosyan}}, \bibinfo {author} {\bibfnamefont {S.}~\bibnamefont
  {Ciliberto}}, \bibinfo {author} {\bibfnamefont {R.}~\bibnamefont
  {Dillenschneider}}, \ and\ \bibinfo {author} {\bibfnamefont {E.}~\bibnamefont
  {Lutz}},\ }\href@noop {} {\bibfield  {journal} {\bibinfo  {journal} {Nature}\
  }\textbf {\bibinfo {volume} {483}},\ \bibinfo {pages} {187} (\bibinfo {year}
  {2012})}\BibitemShut {NoStop}%
\bibitem [{\citenamefont {Lopez-Suarez}\ \emph {et~al.}(2015)\citenamefont
  {Lopez-Suarez}, \citenamefont {Neri},\ and\ \citenamefont
  {Gammaitoni}}]{lopez2015operating}%
  \BibitemOpen
  \bibfield  {author} {\bibinfo {author} {\bibfnamefont {M.}~\bibnamefont
  {Lopez-Suarez}}, \bibinfo {author} {\bibfnamefont {I.}~\bibnamefont {Neri}},
  \ and\ \bibinfo {author} {\bibfnamefont {L.}~\bibnamefont {Gammaitoni}},\
  }in\ \href@noop {} {\emph {\bibinfo {booktitle} {Energy Efficient Electronic
  Systems (E3S), 2015 Fourth Berkeley Symposium on}}}\ (\bibinfo {organization}
  {IEEE},\ \bibinfo {year} {2015})\ pp.\ \bibinfo {pages} {1--2}\BibitemShut
  {NoStop}%
\bibitem [{\citenamefont {Lopez-Suarez}\ \emph {et~al.}(2016)\citenamefont
  {Lopez-Suarez}, \citenamefont {Neri},\ and\ \citenamefont
  {Gammaitoni}}]{lopez2015sub}%
  \BibitemOpen
  \bibfield  {author} {\bibinfo {author} {\bibfnamefont {M.}~\bibnamefont
  {Lopez-Suarez}}, \bibinfo {author} {\bibfnamefont {I.}~\bibnamefont {Neri}},
  \ and\ \bibinfo {author} {\bibfnamefont {L.}~\bibnamefont {Gammaitoni}},\
  }\href@noop {} {\bibfield  {journal} {\bibinfo  {journal} {Nature
  Communications}\ }\textbf {\bibinfo {volume} {7}},\ \bibinfo {pages} {12068}
  (\bibinfo {year} {2016})}\BibitemShut {NoStop}%
\end{thebibliography}%

\end{document}